\documentclass[12pt]{article}
\usepackage{geometry}             % See geometry.pdf to learn the layout options. There are lots.
\usepackage{graphicx}
\usepackage{amssymb}
\usepackage{amsmath}
\usepackage{epstopdf}
\usepackage{cite}

\usepackage[hyperindex=true,
          pdfstartview=FitH,
          bookmarksnumbered=true,
          bookmarksopen=true,
          citecolor=blue,
          linkcolor=blue,
          colorlinks=true,
          pdfborder=001,
          unicode]{hyperref}

\allowdisplaybreaks

\begin{document}

\title{Near horizon symmetry and entropy of black holes in $f(R)$ gravity and conformal gravity }

\author{Jun Li${}^{1}$, Kun Meng${}^{2}$,  Liu Zhao${}^{1}$
\\ ${}^{1}$ School of Physics, Nankai university, Tianjin 300071, China
\\ ${}^{2}$ School of Science, Tianjin Polytechnic University, Tianjin 300387, China
 \\ email: {\it lijun900405@mail.nankai.edu.cn}, {\it mengkun@tjpu.edu.cn} and \\{\it lzhao@nankai.edu.cn}}
\date{\today}                             % Activate to display a given date or no date
\maketitle

\begin{abstract}
We generalize the Noether current method proposed by Majhi and Padmanabhan recently to $f(R)$ gravity and conformal gravity and analyze the near horizon symmetry of black holes in these models. It is shown that the entropy obtained with Cardy's formula agrees with Wald entropy.
\end{abstract}


\section{Introduction}
One of the remarkable properties of black holes is that they are thermal objects
bearing temperature and entropy. The microscopic origin of black hole entropy has
attracted considerable attentions, but a fully satisfactory explanation of
black hole entropy remains to be found. At present, there are at least three
different approaches in explaining black hole entropy, i.e. black hole entropy as
contribution from holographic quantum degrees of freedom (DOF), as quantum
entanglement entropy and as a purely classical geometric entity (Wald entropy).

In the approach explaining black hole entropy as contribution from holographic
quantum DOF, Strominger\cite{Strominger1,Strominger2} has made an important
step forward. He discovered that the asymptotic region of $2+1$ dimensional black
hole spacetimes exhibit conformal symmetry using the Brown-Henneaux's canonical
approach\cite{BrownHenneaux}, and the black hole entropy calculated using Cardy's
formula \cite{Cardy} agrees exactly with the Bekenstein-Hawking entropy.
Carlip \cite{Carlip} found that the near horizon region of black holes also exhibits
conformal symmetry, and the black hole entropy can still be calculated using
Cardy's formula. Later on, Strominger et al studied the near horizon conformal
symmetry of axisymmetric black holes which is now known as Kerr/CFT
\cite{Strominger3}.

Recently, Majhi and Padmanabhan proposed an alternative method (MP approach for
short) to study the asymptotic symmetry of the black hole solution \cite{Majhi,Majhi:2012nq}.
They considered the Noether currents related only to the diffeomorphism
invariance of the {\em boundary action}, which is different from the bulk ones
previously considered. By solving the Killing vector fields that generate the
boundary Noether currents and studying the anomaly of corresponding conformal
algebra, the black hole entropy computed via Cardy's formula matches
Bekenstein-Hawking entropy precisely without shifting the zero-mode energy
\cite{Carlip,Park}. Several works have been done generalizing MP approach
to higher derivative gravity \cite{SJZ,Yi,Hirochi} or gravity with matter source
\cite{Meng}. For further applications of the boundary diffeomophims in the context of
the entropies of gravity with one loop effective modified action, of the 
conformally transformed black hole solution as well as for time dependent black holes, see
Refs. \cite{Majhi:2013lba,Majhi:2014lka,Majhi:2015tpa}.

In this paper we will consider the role of MP approach in
two interesting models of higher curvature gravity, i.e. $f(R)$ gravity and conformal 
gravity. As a modified gravity model for interpreting cosmic acceleration, $f(R)$
gravity has been proven to be able to mimic the whole cosmological history, from
inflation to later accelerating expansion era. There have been many studies on
the application of $f(R)$ to cosmology and gravitation\cite{Capozziello}.
In four dimensions, Weyl-squared conformal gravity is the unique conformally
invariant gravity that is polynomial in curvature. It has an important feature
that all metrics which are conformal transformations of solutions of Einstein
gravity are automatically solutions of conformal gravity. Recently, Maldacena
proved that, by imposing proper boundary conditions, Einstein gravity can emerge
from conformal gravity in four dimensions. Thus, in order to understand the
microscopic origin of black hole entropy and make comparison between different
approaches, we would like to study the near horizon symmetries of static black
holes of the above two models using MP approach.

The paper is organized as follows. In Section 2, we give a brief review of the
MP approach. In Section 3, we compute the conserved charges owing to the
diffeomorphism invariance of the boundary action of $f(R)$ gravity and obtain
the related Virasoro algebra. The entropy computed via Cardy's formula matches
the Wald entropy exactly. In Section 4, we compute the conserved charges, construct the
corresponding Virasoro algebra and work out the entropy with Cardy's formula in 
conformal gravity. We conclude our results in Section 5.

\section{Noether currents and charges from boundary action}

In this section, we briefly review the boundary Noether current method in Einstein gravity
proposed by Majhi and Padmanabhan recently\cite{Majhi}.
We will start with the boundary term of a general action which is written as a
total divergence
\begin{align}
\sqrt{g}L=\sqrt{g}\nabla_a A^a.\label{Lag}
\end{align}
Under a diffeomorphism $x^a\rightarrow x^a+\xi^a(x)$, the variation of the left hand side of (\ref{Lag}) is the Lie derivative of a scalar
\begin{equation}
\delta_{\xi}\left(\sqrt{g}L\right)
\equiv\mathcal{L}_{\xi}\left(\sqrt{g}L\right)
=\sqrt{g}\nabla_a\left(L\xi^a\right),\label{varileft}
\end{equation}
where we have used $\delta\sqrt{g}=\frac{1}{2}\sqrt{g}g^{ab}\delta g_{ab}$ and $\delta_{\xi}g_{ab}=\nabla_a \xi_b+\nabla_b \xi_a$. The variation of the right hand side of is
\begin{align}
&\delta_{\xi}\left(\sqrt{g}\nabla_aA^a\right)
=\mathcal{L}_{\xi}\left[\partial_a
\left(\sqrt{g}A^a\right)\right]\nonumber\\
&\quad=\sqrt{g}\nabla_a\left[\nabla_b\left(A^a\xi^b\right)
-A^b\nabla_b\xi^a\right], \label{varright}
\end{align}
where $\nabla_aA^a=\frac{1}{\sqrt{g}}\partial_a
\left(\sqrt{g}A^a\right)$ is used.

After equating Eq.(\ref{varileft}) and Eq.(\ref{varright}), we can read off the  Noether current
\begin{align}
  J^a\left[\xi\right]=L\xi^a-\nabla_b
  \left(A^a\xi^b\right)+A^b\nabla_b\xi^a.\label{current}
\end{align}
which obeys $\nabla_aJ^a=0$. Replacing $L$ in (\ref{current}) with (\ref{Lag}), the Noether current can be rewritten as
\begin{align}
  J^a\left[\xi\right]=\nabla_bJ^{a b}\left[\xi\right]=\nabla_b\left[\xi^a A^b-\xi^b A^a\right].\label{current2}
\end{align}
Given the Noether current, the conserved charge is defined as
\begin{align}
  Q\left[\xi\right]&=\int_\Sigma \mathrm{d}\Sigma_a J^a
  =\frac{1}{2}\int_{\partial\Sigma}\sqrt{h}\,\mathrm{d}
  \Sigma_{ab}\,J^{ab},\label{charge}
\end{align}
where $\Sigma$ is a timelike hypersurface with unit normal vector
$M^a$. We have used Stokes' theorem in the second equality.
$\mathrm{d}\Sigma_{ab}=-\mathrm{d}^{(d-1)}x
\left(N_aM_b-N_bM_a\right)$ is the area element on the $(d-1)$-dimensional
hypersurface $\partial\Sigma$, and $N^a$
is the spacelike unit normal vector.

Finally, the Lie bracket of the conserved charges is defined as \cite{Majhi:2011ws}
\begin{align}
  &\left[Q_1,Q_2\right]\equiv\frac{1}{2}
  \left(\delta_{\xi_1}Q[\xi_a]-\delta_{\xi_a}
  Q[\xi_1]\right)\nonumber\\
  &\qquad=\frac{1}{2}\int_{\partial\Sigma}\sqrt{h}
  \mathrm{d}\Sigma_{ab}\bigg(\xi_2^aJ^b[\xi_1]
  -\xi_1^aJ^b[\xi_2]\bigg),\label{commutator}
\end{align}
where $\delta_{\xi_1}Q[\xi_2]=\int_{\Sigma}
\mathrm{d}\Sigma_a\,\mathcal{L}_{\xi_1}\left(
\sqrt{g}J^a[\xi_2]\right)$. For Einstein gravity this commutator gives rise to
the Virasoro algebra \cite{Majhi} and the black hole entropy follows directly
by employing Cardy's formula.

\section{$f(R)$ gravity}

In this section, we aim to generalize MP approach to $f(R)$ gravity in
$(d+1)$ spacetime dimensions. The bulk
action reads \cite{Guarnizo}
\begin{align}
I=\frac{1}{16\pi G}\int\mathrm{d}^{d+1}x\sqrt{-g}f(R),\label{fRbulk}
\end{align}
where $f(R)$ is a polynomial of the Ricci scalar $R$, in which a term
proportional to $R$ (Einstein-Hilbert term) and a constant term (the cosmological
constant term) may be present.

The boundary action which cancels the total divergence term arising from the
variation of the bulk action (\ref{fRbulk}) %\cite{Guarnizo:2010xr}
is
\begin{align}
I_B=\frac{1}{8\pi G}\int \mathrm{d}^{d}x\sqrt{\sigma}f'(R)K,\label{BTfR}
\end{align}
where $K$ is the trace of the extrinsic curvature and the prime denotes
the derivative with respect to the Ricci scalar $R$.

Now let's consider a static black hole solution of $f(R)$ gravity
\begin{align}
\mathrm{d}s^2=-f(r)\mathrm{d}t^2+\frac{\mathrm{d}r^2}{f(r)}
+r^2\Omega_{ij}(x)\mathrm{d}x^i \mathrm{d}x^j
\end{align}
where $\Omega_{ij}(x)$ is the metric on a $(d-1)$-dimensional unit sphere
and we denote $h_{ij}=r^2\Omega_{ij}(x), h=\mathrm{det}\,h_{ij}$.
The unit normal vectors in this spacetime can be chosen as
\begin{align}
N^a=\big(0,\sqrt{f(r)},0,\cdots,0\big),
M^a=\bigg(\frac{1}{\sqrt{f(r)}},0,\cdots,0\bigg).
\end{align}
Assume that the event horizon of the black hole lies at $r=r_h$ with $f(r_h)=0$.
In order to describe the near-horizon geometry, let's introduce a local Rindler
coordinate $\rho$ via $r=r_h+\rho$, in terms of which the metric becomes
\begin{align}
\mathrm{d}s^2=-f(r_h+\rho)\mathrm{d}t^2+\frac{\mathrm{d}\rho^2}{f(r_h+\rho)}
+(r_h+\rho)^2\Omega_{ij}(x)\mathrm{d}x^i \mathrm{d}x^j.\label{metric}
\end{align}
In the near horizon region, $\rho\to0$, the function $f(r_h+\rho)$ can be
expanded as
$f(r_h+\rho)=2\kappa\rho+\frac{1}{2}f''(r_h)\rho^2+\cdot\cdot\cdot$, with
$\kappa=\frac{f'(r_h)}{2}$ being the surface gravity. In order to get the proper
vector field $\xi^\mu$ that generates the boundary diffeomorphism invariance,
we transform this metric to the Bondi-like form and then transform back
after solving the Killing equations . Under the transformation
\begin{align}
    \mathrm{d}u=\mathrm{d}t-\frac{\mathrm{d}\rho}{f(r_h+\rho)},
\end{align}
the metric (\ref{metric}) can be rewritten as
\begin{align}
    \mathrm{d}s^2=-f(r_h+\rho)\mathrm{d}u^2-2\mathrm{d}u\mathrm{d}\rho
    +(r_h+\rho)^2\Omega_{ij}(x)\mathrm{d}x^i \mathrm{d}x^j.
\end{align}
Requiring invariance of the horizon structure under infinitesimal
boundary diffeomorphisms yields the following Killing equations,
\begin{align}
\mathcal{L}_\xi g_{\rho\rho}&=-2\partial_\rho\xi^u=0,\nonumber\\
\mathcal{L}_\xi g_{u\rho}&=-f(r_h+\rho)\partial_\rho\xi^u-\partial_\rho\xi^\rho-
\partial_u\xi^u=0.
\end{align}
Solving the above Killing equations we get
\begin{align}
\xi^u=F(u,x),\ \ \ \xi^\rho=-\rho\partial_uF(u,x),\label{killing}
\end{align}
where $x$ denotes the coordinates on the unit $(d-1)$-sphere and $F(u,x)$ is
an arbitrary function of the arguments. All other
components of $\xi^a$ vanish. The condition $\mathcal{L}_\xi g_{uu}= 0$ is
satisfied automatically in the near-horizon limit given the Killing vectors (\ref{killing}).

Transforming back to the original coordinates ($t,\rho$), the Killing vector
fields now take the form
\begin{align}
\xi^t&=T-\frac{\rho}{f(\rho+r_h)}\partial_tT,\nonumber\\
\xi^\rho&=-\rho\partial_tT,
\end{align}
where $T$ is the same arbitrary function $F(u,x)$ but with coordinate $u$ changed
into $t$ and $\rho$. For each choice of $T$
there is a corresponding Killing vector field $\xi^a$.
$T$ can be expanded in terms of a set of basis functions
\begin{align}
    T=\sum A_m T_m
\end{align}
where the basis function $T_m$ should be chosen properly in order that the Diff $S^1$ algebra is satisfied
\begin{align}
    i\{\xi_m,\xi_n\}^a=(m-n)\xi^a_{m+n}.
\end{align}
Here $\{,\}$ is the Lie bracket.  A standard choice of $T_m$ is
\begin{align}
T_m=\frac{1}{\alpha}\exp[im(\alpha t+g(\rho)+p\cdot x)],\label{mode}
\end{align}
where $\alpha$ is a constant, $p$ is an integer, $g(\rho)$ is a regular function
on the horizon. For $T$ to be real, the coefficients $A_m$ must obey
$A_m^\ast=A_{-m}$.

Now we have all the ingredients to calculate the conserved charges and the
commutators between them. For $f(R)$ gravity, we can read off $A^a=N^af'(R)K$
from the boundary action (\ref{BTfR}). Inserting this $A^a$ into Eq.
(\ref{current2}) the Noether potential follows,
\begin{align}
J^{ab}[\xi]=\frac{1}{8\pi G}f'(R)K[\xi^aN^b-\xi^bN^a].\label{potential}
\end{align}
Recall that $J^a=\nabla_bJ^{ab}$. Using this and substituting in (\ref{charge}) and Eq.(\ref{commutator}) and then taking the limit $\rho\to 0$, we get
\begin{align}
Q&=\frac{1}{8\pi G}\int d^{d-1}x\sqrt{h}f'(R)|_{\rho=0}
\bigg(\kappa T-\frac{1}{2}\partial_tT\bigg),\label{charge1}\\
[Q_1,Q_2]&=\frac{1}{8\pi G}\int d^{d-1}x\sqrt{h}f'(R)|_{\rho=0}
\bigg[\kappa\big(T_1\partial_tT_2-T_2\partial_tT_1\big)-
\frac{1}{2}\bigg(T_1\partial_t^2T_2-T_2\partial_t^2T_1\bigg)\nonumber\\
&+\frac{1}{4\kappa}\big(\partial T_1\partial_t^2T_2-\partial T_2\partial_t^2T_1
\big)\bigg].\label{commutator1}
\end{align}
Comparing this result to the case of pure Einstein gravity
\cite{Majhi}, one finds that an additional factor $f'(R)|_{\rho=0}$ appears.
Choosing the standard form (\ref{mode}) of $T_m$ and making the integrations,
we get the final form of the modes $Q_m$ and the commutators
\begin{align}
Q_m&=\frac{1}{8\pi G}f'(R)|_{\rho=0}\frac{\kappa A}{\alpha}\delta_{m,0},\\
[Q_m,Q_n]&=\frac{1}{8\pi G}f'(R)|_{\rho=0}\bigg[-\frac{i\kappa A}{\alpha}(m-n)
\delta_{m+n,0}-im^3\frac{\alpha A}{2\kappa}\delta_{m+n,0}\bigg].
\end{align}
The last commutator reminds us the form of the Virasoro algebra, from which we can read off the central charge $C$ and the zero mode $Q_0$,
\begin{align}
    \frac{C}{12}=\frac{\alpha A}{2\kappa}\frac{1}{8\pi G}f'(R)|_{\rho=0},
\end{align}
\begin{align}
    Q_0=\frac{A\kappa}{\alpha}\frac{1}{8\pi G}f'(R)|_{\rho=0}.
\end{align}
So, using Cardy's formula, we get the entropy of the black hole
\begin{align}
    S=2\pi\sqrt{\frac{CQ_0}{6}}=\frac{A}{4G}f'(R)|_{\rho=0}.
\end{align}
This is exactly the Wald entropy of static black hole in $f(R)$ gravity.

\section{Conformal gravity}

In this section, we apply the MP approach to the static black hole in
four-dimensional conformal gravity.

The boundary action of conformal gravity \cite{Lu,Grumiller:2013mxa} can be
written as
\begin{align}
I_B=\frac{\alpha_{c}}{4\pi}\int \mathrm{d}^3x\sqrt{\sigma}C^{abcd}n_a n_c\nabla_b n_d,\label{CG}
\end{align}
where $n^a$ is the unit normal vector of the boundary of the spacetime. If we
consider the metric of static black hole with the same form as in (\ref{metric}),
then $n^a=\bigg(0,\frac{1}{\sqrt{f(r)}},0,0\bigg)$. From (\ref{CG}) it's very
easy to read off the boundary Lagrangian density
\begin{align}
    \mathcal{L}_B=\frac{\alpha_{c}}{4\pi} C^{abcd}n_a n_c\nabla_b n_d,
\end{align}
So, we can take $A^a=n^a\mathcal{L}_B$ and insert this object
into Eqs.(\ref{current2}--\ref{commutator}). After taking the
limit $\rho\to 0$, we get
\begin{align}
Q&=\frac{\alpha_{c}}{4\pi}\frac{2+2r_hf'(r_h)-r_h^2f''(r_h)}{6r_h^2}
\int \mathrm{d}^{2}x\sqrt{h}\bigg(\kappa T-\frac{1}{2}\partial_tT\bigg),\\
[Q_1,Q_2]&=\frac{\alpha_{c}}{4\pi}\frac{2+2r_hf'(r_h)-r_h^2f''(r_h)}{6r_h^2}\int \mathrm{d}^{2}x\sqrt{h}\bigg[\kappa(T_1\partial_tT_2-T_2\partial_tT_1)\nonumber\\
&-\frac{1}{2}(T_1\partial_t^2T_2-T_2\partial_t^2T_1)
+\frac{1}{4\kappa}(\partial T_1\partial_t^2T_2-\partial T_2\partial_t^2T_1)\bigg]
\end{align}

Choosing the standard form (\ref{mode}) for $T_m$ and integrating out the above
equations, we get the final form of the modes $Q_m$ and the commutators
\begin{align}
Q_m&=\frac{\alpha_{c}}{4\pi}\frac{2+2r_hf'(r_h)-r_h^2f''(r_h)}{6r_h^2}
\frac{\kappa A}{\alpha}\delta_{m,0},\\
[Q_m,Q_n]&=\frac{\alpha_{c}}{4\pi}\frac{2+2r_hf'(r_h)-r_h^2f''(r_h)}{6r_h^2}
    \bigg[-\frac{-i\kappa A}{\alpha}(m-n)\delta_{m+n,0}
    -im^3\frac{\alpha A}{2\kappa}\delta_{m+n,0}\bigg].
\end{align}
Therefore we recognize the Virasoro central charge $C$ and the zero mode $Q_0$
\begin{align}
\frac{C}{12}&=\frac{\alpha A}{2\kappa}\frac{\alpha_{c}}{4\pi}\frac{2+2r_hf'(r_h)
-r_h^2f''(r_h)}{6r_h^2},\\
Q_0&=\frac{\kappa A}{\alpha}\frac{\alpha_{c}}{4\pi}\frac{2+2r_hf'(r_h)
-r_h^2f''(r_h)}{6r_h^2}.
\end{align}
Employing Cardy's formula, we  get the entropy of the black hole
\begin{align}
    S=2\pi\sqrt{\frac{CQ_0}{6}}=\frac{\alpha_{c}}{8}\bigg(\frac{4}{3r_h^2}+
    \frac{4f'(r)}{3r_h}-\frac{2f''(r)}{3}\bigg)A.
\end{align}
This agrees exactly with the Wald entropy of the static black hole
\cite{Lu,Bodendorfer:2013wga,Sinha:2010ai,Liu:2012xn}
\begin{align}
    S=-\frac{\alpha_{c}}{8}\int C_{abcd}\epsilon^{ab}\epsilon^{cd}
    \mathrm{d}\Sigma=
    \frac{\alpha_{c}}{8}\bigg(\frac{4}{3r_h^2}+\frac{4f'(r)}{3r_h}
    -\frac{2f''(r)}{3}\bigg)A.
\end{align}

\section{Conclusion}
It was known in the literature that for higher curvature extended gravity models,
the Beckenstein-Hawking formula for black hole entropy is not valid.
Instead, the Wald entropy is a universal formula for all gravity models.
However, the Wald entropy is constructed purely from geometric considerations
and thus seems to be a purely classical object. This intuition contradicts
with the expectation that black hole entropy may arise from microscopic
quantum DOF of gravity. On the other hand, the MP approach, which makes use of
the off-shell boundary conserved charges and the Cardy's formula, seems to relate
both geometric and quantum considerations and hence feels more natural
for interpreting black hole entropy.

In this paper, we calculated the black hole entropy in both $f(R)$ gravity and
conformal gravity using the MP approach. The results in both cases agree with
the Wald entropy. This provides further evidence for the universality of the
Wald entropy. Besides, in the MP approach, we made use of only the staticity and the
spherical symmetry of the metric but not of the explicit form of
the metric function $f(r)$. This makes the construction equally well applicable
to static spherically symmetric black holes in other extended theories of
gravity.

The importance of the boundary action is not limited within the context of variational problems for gravitational theories and the boundary diffeomorphism invariance. 
Recently, Majhi promoted the boundary action to the calculation of black hole entropy using the entropy function method in general relativity \cite{Majhi:2015pra} and the result agrees
with the boundary diffeomophism approach. It is desirable to see whether the boundary entropic function method is also applicable to other models of gravity.

\section*{Acknowledgement}

The work of KM is supported by the
National Natural Science Foundation of China (NSFC) under grant No.11447153.

\providecommand{\href}[2]{#2}\begingroup%\raggedright
\footnotesize\itemsep=0pt
\providecommand{\eprint}[2][]{\href{http://arxiv.org/abs/#2}{arXiv:#2}}

\end{document}